\documentclass[twocolumn,showpacs,preprintnumbers,amsmath,amssymb]{revtex4}
\usepackage{graphicx}
\usepackage{dcolumn}
\usepackage{bm}

\begin{document}

\title{One-shot quantum measurement using a hysteretic DC-SQUID}

\author{O. Buisson$^{1}$, F. Balestro$^{1}$, J.P. Pekola$^{1,2}$, and F.W.J. Hekking$^{3}$}
\affiliation{$^{1}$Centre de Recherches sur les Tr\`es Basses
Temp\'eratures, laboratoire associ\'e \`a l'Universit\'e Joseph
Fourier, C.N.R.S., BP 166, 38042 Grenoble-cedex 9, France\\
$^{2}$Low Temperature Laboratory, Helsinki University of
Technology, P.O. Box 2200, 02015 HUT, Finland\\
$^{3}$Laboratoire de Physique et Mod\'elisation des Milieux
Condens\'es, CNRS \& Universit\'e Joseph Fourier, BP 166, 38042
Grenoble-cedex 9, France}

\begin{abstract}
We propose a single shot quantum measurement to determine the
state of a Josephson charge quantum bit (qubit). The qubit is a
Cooper pair box (CPB) and the measuring device is a two junction
superconducting quantum interference device (dc-SQUID). This
coupled system exhibits a close analogy with a Rydberg atom in a
high Q cavity, except that in the present device we benefit from
the additional feature of escape from the supercurrent state by
macroscopic quantum tunneling, which provides the final read-out.
We test the feasibility of our idea against realistic experimental
circuit parameters and by analyzing the phase fluctuations of the
qubit.
\end{abstract}


\maketitle

A Cooper pair box is a controllable macroscopic two level
system~\cite{Bouchiat98,Nakamura99}, which is considered as a potential qubit in the
context of quantum computing \cite{Makhlin01}. Coherent Rabi oscillations have been
observed in a CPB, using ultrafast pulses~\cite{Nakamura99}. Recently, in a
superconducting device named "quantronium", coherent oscillations were observed using
microwave pulses \cite{Vion02}. The observed oscillations lived for almost a
microsecond, making superconducting circuits promising for realizing quantum gates.

Despite this progress, quantum measurement on a CPB still remains
a challenge. The charge read-out circuits, like the one
in~\cite{Nakamura99} or those using single electron tunneling
transistors (see, {\em e.g.}, \cite{Aassime01}) are far from ideal
due to decoherence induced by the unavoidable background charge
fluctuations and by backaction during the measurement, {\em e.g.},
in form of shot noise. The former problem can be avoided to a
large extent if the qubit is operated at the degeneracy point of
the two level system: at this optimum point linear coupling to
charge fluctuations is absent. In the quantronium experiment, both
problems were eliminated successfully by measuring the two quantum
states of a Cooper pair transistor (CPT) at the optimum point by a
hysteretic Josephson junction. This junction, working in its
classical regime, measured the persistent current of the two
quantum states of the CPT. This method cannot, however, be used to
measure the quantum state of a CPB. Moreover, the measurement did
not resolve the quantum state in one shot.

In this letter we propose a new quantum measurement procedure based on the
entanglement between two quantum systems: a CPB coupled to a superconducting
resonator. The dynamics of this coupled system has been theoretically investigated
recently~\cite{Buisson00,Hekking01,Marquardt00,Al-Saidi02}. Here we show that at the
degeneracy the two quantum states of the CPB can be resolved in one shot. The read-out
device is a current-biased dc-SQUID (the resonator) and it is controlled by adiabatic
pulses of flux. Very high sensitivity and fast measurement can be reached with this
method. A dc-SQUID is preferred over a Josephson junction because manipulations using
external flux do not suffer from time limitations. In a Josephson junction, nanosecond
pulses of bias current cannot be applied because low pass filters are necessary to
exclude high-frequency noise~\cite{Martinis87}.

The circuit of a Cooper pair box coupled to a current-biased dc-SQUID was inspired by
experiments on a Rydberg atom in a high Q cavity \cite{Brune96}. For a bias current
$I$ well below the critical current $I_{\rm c}$, the CPB and the SQUID play the roles
of the Rydberg atom and the high Q cavity, respectively. For example, Rabi
oscillations are predicted to occur as a result of spontaneous emission and
re-absorption by the CPB of a single oscillation quantum in the
SQUID~\cite{Buisson00}. The device can be considered as a two level system coupled to
a harmonic oscillator. However, for a bias current very close to, but below the
critical current, macroscopic quantum tunneling (MQT) in the SQUID can occur. This
quantum escape phenomenon has no equivalent in high Q cavity experiments, and it
introduces a new element into the dynamics of the system. We will describe how we can
benefit from MQT in order to do a very fast one-shot quantum measurement on the CPB.
We will discuss the performance of this measurement and its back-action on the CPB.

\begin{figure}
\resizebox{.38\textwidth}{!}{\includegraphics{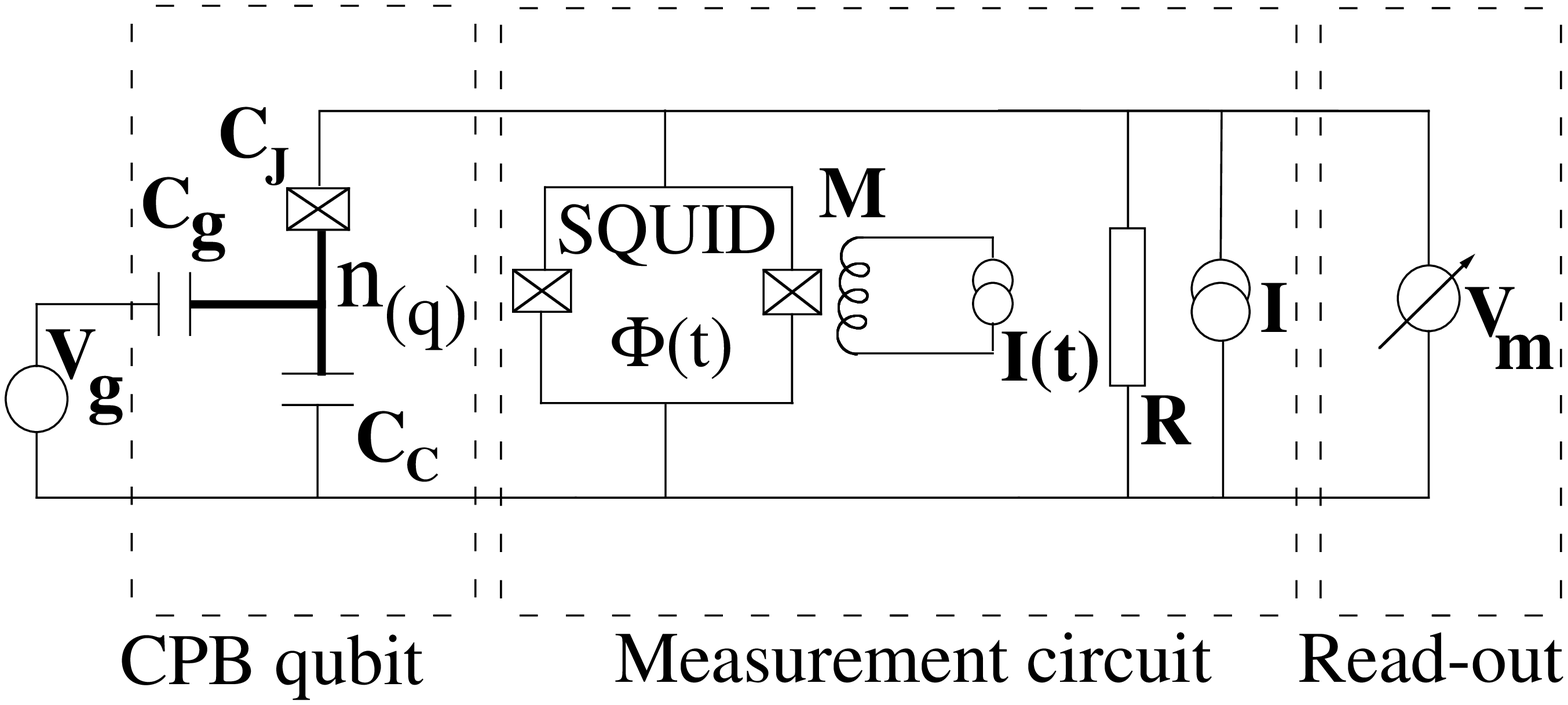}}
\caption{\label{device} The CPB-qubit is coupled to a measuring
SQUID through the capacitance $C_{c}$. $V_{g}$ controlled the CPB
states, $I$ and $I(t)$ the measurement procedure. The voltmeter
$V_{m}$ performs the read-out.}
\end{figure}

The superconducting quantum circuit that we consider is shown in
Fig.~\ref{device}. It contains three parts which we describe in
detail below: a Cooper pair box, a hysteretic dc-SQUID, and a
coupling capacitance between the two. The circuit is connected to
the external classical circuits by three different couplings: a
resistance $R$ in parallel to a DC current source and voltmeter, a
mutual inductance to a source of flux modulating current pulses,
and a gate capacitance to a dc and pulse voltage source.

The {\em Cooper pair box} consists of a small superconducting island, coupled to a
gate-voltage $V_{\rm g}$ by a gate capacitance $C_{\rm g}$. The island is furthermore
connected capacitively to a superconducting electrode via a Josephson junction with a
capacitance $C_{\rm J}$ and Josephson energy $E_{\rm J}$. We are interested in the
limit of small Josephson energy, $E_{\rm J} \ll E_{\rm C,J}$, where $E_{\rm
C,J}=2e^2/C_{\rm J,eff}$ is the elementary charging energy of the box with $C_{\rm
J,eff} = C_{\rm J} +[1/C_{\rm S} + 1/(C_{\rm c}+C_{\rm g})]^{-1}$; $C_{S}$ is the
SQUID capacitance and $C_{\rm c}$ the coupling capacitance. Let us introduce the basis
of charge states $|n_{\rm (q)}\rangle$, where $n_{\rm (q)}$ corresponds to the number
of excess Cooper pairs on the island. If the dimensionless gate charge $N_{\rm g} =
-C_{\rm g}V_{\rm g}/(2e)$ is close to $ 1/2$, the charge states $|0 _{\rm (q)}\rangle$
and $|1 _{\rm (q)}\rangle$ are almost degenerate and the relevant eigenstates of the
CPB are superpositions of these charge states. Specifically, the ground state and the
first excited state are, respectively: $|-\rangle = ( |0_{\rm (q)}\rangle + |1_{\rm
(q)}\rangle)/\sqrt{2}$ and $|+\rangle = ( |0_{\rm (q)}\rangle - |1_{\rm (q)}\rangle)
/\sqrt{2}$. The corresponding eigenenergies are $E_{\mp} = E_{\rm C,J} \mp \frac{1}{2}
E_{\rm J}$. We see that the CPB effectively behaves as a quantum-mechanical two-level
system~\cite{Makhlin01}.

The {\em hysteretic dc-SQUID} consists of a superconducting loop
with two underdamped Josephson junctions which both have ideally
the same critical current $I_{0}$ and capacitance $C_{\rm S}/2$.
We neglect the loop inductance and effects due to asymmetry in the
SQUID. The damping of the two junctions is also neglected now, but
it will be discussed at the end. With these approximations, the
SQUID equation of motion is similar to that of a single Josephson
junction, describing a particle of mass $m=C_{\rm S,eff}(\Phi
_{0}/(2\pi))^2$ in a tilted washboard potential $
U(\varphi)=E_{\rm S} [-I \varphi /I_{\rm c}-\cos(\varphi)]$, where
$\varphi$ is the phase difference of the SQUID, $I$ the bias
current through the SQUID, $\Phi _{0} = h/2e$ the superconducting
flux quantum, $\Phi _{\rm DC}$ the external flux, and $I_{\rm c}=2
\left\vert{\cos(\pi \Phi _{\rm DC}/\Phi _{0})}\right\vert I_{0}$, $E_{\rm S}=I_{\rm
c}\Phi _{0}/(2\pi)$ and $C_{\rm S,eff}=C_{\rm S}+[1/C_{\rm
J}+1/(C_{\rm c}+C_{\rm g})]^{-1}$ the effective critical current,
Josephson energy and capacitance of the SQUID, respectively.

For values of bias current not too far below the critical current,
the potential $U(\varphi)$ can be well approximated by a cubic
potential. The quantum dynamics of the SQUID using the reduced
momentum and position operators $\hat{P}=(1/\sqrt{m \hbar \omega
_{\rm p}}) P$ and $\hat{X}=(\sqrt{m \omega _{\rm p}/\hbar}) X$,
respectively, is then described by $
\hat{H}_{0}=\frac{1}{2}\hbar\omega _{p} (\hat{P}^2+\hat{X}^2)+
\sigma (I)\hbar\omega _{\rm p}\hat{X}^3$, where $X=\varphi$ is the
phase difference, $P$ its conjugate operator and $\omega _{\rm
p}=[2\pi I_{c}/(\Phi_{0} C_{\rm S,eff})]^{1/2}[2(1-I/I_{\rm
c})]^{1/4}$ the effective plasma frequency of the SQUID. The
parameter $\sigma (I) =-1/(6a) [2(1-I/I_{\rm c})]^{-3/4}$, where
$a=\hbar^{-1/2}[\Phi_{0}/(2\pi)]^{3/4} (C_{\rm S,eff}I_{\rm
c})^{1/4}$, gives the relative magnitude of the cubic term as
compared to that of the harmonic oscillator term.

For values of $I$ below $I_{\rm c}$ with $\sigma (I)\ll 1$, the potential barrier is
high compared to $\hbar\omega _{\rm p}$, and the cubic term in $\hat{H}_{0}$ can be
neglected. Many low-lying states are found near the minimum of the quadratic
potential. The broadening of these states due to tunneling can be ignored. Hence these
states are well approximated by harmonic oscillator eigenstates, denoted by
$|0\rangle, |1\rangle, |2\rangle, ...$, corresponding to the presence of 0, 1, 2, ...
oscillation quanta in the SQUID, respectively. Thus, at low enough bias current, the
SQUID behaves as a superconducting quantum resonator. Since in this limit the phase is
localized in a well defined minimum of the potential $U$, the time-averaged voltage
$V_{\rm m}$ across the SQUID remains zero.

If the bias current is increased such that $I \alt I_{\rm c}$,
$\sigma (I)$ is no longer negligible and the cubic term affects
the SQUID dynamics. The barrier height, given by $\Delta
U=4\sqrt{2}/3 E_{\rm S}(1-I/I_{\rm c})^{3/2}$, and $\omega _{\rm
p}$ decrease and vanish at the critical current. The number of
localized states in a given well decreases. Moreover, the
remaining energy levels broaden due to quantum tunneling from the
metastable wells of the potential $U$. The broadening of the
ground state energy for low damping is given by the tunneling rate
$\Gamma _0 = \omega _{\rm p} 6\sqrt{6/\pi}\sqrt{\Delta U/\hbar
\omega_{\rm p}} \exp(-36\Delta U/(5\hbar \omega_{\rm p}))$. Since
the excited states $|n\rangle$ are located closer to the top of
the barrier, the tunneling rates and hence the broadening of these
states increase with increasing energy of the level as $\Gamma _n
\sim (\exp(-36/5))^n \Gamma_{0}\,$. Specifically, the tunneling
rates from states $|0\rangle$ and $|1\rangle$ of the SQUID are
different by approximately a factor thousand
\cite{Weiss99,Larkin92}. After a tunneling event, the phase of the
SQUID is no longer localized and the time-averaged voltage $V_{\rm
m}$ across the SQUID is finite.

The magnetic flux through the SQUID affects its critical current
$I_{\rm c}$, hence $\omega _{\rm p}$, $\sigma (I)$ and therefore
$\hat{H}_{0}$ depend on the flux. For small, time-dependent
variations of flux, $\delta \Phi(t) \ll \Phi _{0}$, and for SQUID
parameters such that $a \gg 1$, the total time-dependent
Hamiltonian is given by $ \hat{H}(t)=\hat{H}_{0}-\lambda
(I,\Phi_{\rm DC}) (\delta \Phi(t)/\Phi _{0}) \hbar\omega_{p}
\hat{X} $, where $\lambda (I,\Phi _{\rm DC}) =a[2(1-I/I_{\rm
c})]^{-3/8} I/I_{\rm c} \pi \tan(\pi \Phi _{\rm DC}/\Phi _{0})$.
Below we consider the effect of a time-dependent perturbation
$\delta \Phi(t)$ on the dynamics of the coupled system.

The {\em coupling capacitance} $C_{\rm c}$ plays a crucial role in
the circuit of Fig.~\ref{device} since it couples the qubit and
the SQUID to each other. Physically, $C_{\rm c}$ couples the
charge $n_{(\rm q)}-N_{\rm g}$ on the CPB to the charge on the
SQUID. The coupling Hamiltonian can be written as $ \hat{H}_{\rm
c} = - i E_{\rm coupl}  (n_{(\rm q)}) \hat{P}$. Here we introduced
the characteristic coupling energy $E_{\rm coupl} = \sqrt{\hbar
\omega _{\rm p}/E_{\rm C,S}} E_{\rm C,c}/4$ where $E_{\rm C,c} =2
e^2/C_{\rm c,eff}$ and $E_{\rm C,S}=2 e^2/C_{\rm S,eff}$ with
$C_{\rm c,eff} = C_{\rm c}+C_{\rm g}+(1/C_{\rm J}+1/C_{\rm
S})^{-1}[(C_{\rm J}+C_{\rm S})/(C_{\rm c}+C_{\rm g})]/2$. This
coupling energy leads to full entanglement between the states of
the CPB and the SQUID at the resonance condition $E_{\rm J}=\hbar
\omega _{\rm p}$~\cite{Buisson00}.

\begin{figure}
\resizebox{.38\textwidth}{!}{\includegraphics{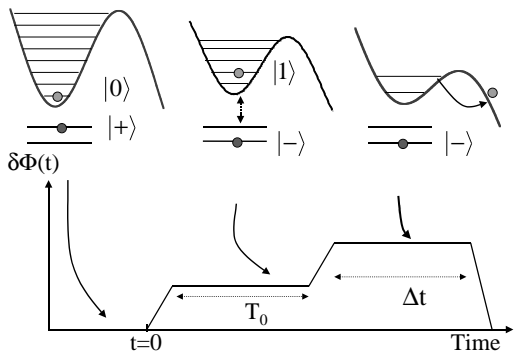}}
\caption{\label{procedure}The quantum measurement procedure is
illustrated for an initially state $|+\rangle$.}
\end{figure}

Having detailed the superconducting quantum circuit, we will now
describe the measurement procedure. Suppose that, at time $t=0$,
the CPB is in a coherent superposition $\alpha |-\rangle +\beta
|+\rangle$, as a result of quantum operations performed at times
$t<0$ using the gate-voltage $V_{\rm g}$~\cite{Nakamura99}. We
propose a way to measure the probability $|\beta|^2$. The
measurement procedure is depicted schematically in
Fig.~\ref{procedure} and consists of three successive step-like
variations of flux through the SQUID. The steps are not sharp, but
have a finite rise and fall time $\delta t$. The flux steps must
be {\em adiabatic} in terms of the dynamics of the CPB, $\delta
t\gg h/E_{\rm J}$, and of the SQUID, $\delta t\gg
2\pi/\omega_{p}$. But they must be {\em instantaneous} in terms of
the coupling dynamics, $\delta t \ll h/E_{\rm coupl}$. The first
step at $t=0$ puts the SQUID into resonance with the CPB during a
time $T_{0}$. The second step at $T_{0}$ drives the SQUID close to
its critical current during a time $\Delta t$. The last step sets
the SQUID far below the critical current.

In more detail, at $t<0$, during the quantum manipulation of the
CPB, the SQUID must be decoupled. This condition is achieved if
$(\hbar \omega _{\rm p}-E_{\rm J})\gg E_{\rm coupl}$, {\em i.e.},
off-resonance. Thus, in leading order, the eigenstate of the
entire system is a product of the eigenstates of the qubit and the
SQUID, in spite of the presence of the coupling capacitance. At
$t=0$, suppose the SQUID in its ground state $|0\rangle$. The
quantum state of the entire system is then
$|\psi(t=0)\rangle=[\alpha |-\rangle +\beta |+\rangle]\otimes
|0\rangle$.

The flux step applied at $t=0$ reduces the effective critical
current and the resonant condition $E_{\rm J}=\hbar \omega _{\rm
p}$ is satisfied for a time $T_{0}$. The state $|-\rangle \otimes
|0\rangle$ is still stationary at this resonance since $|E_{\rm
J}+\hbar \omega _{\rm p}| \gg E_{\rm coupl}$. But $|+\rangle
\otimes |0\rangle$ is no longer an eigenstate: the state of the
coupled system oscillates in time between $|+\rangle \otimes
|0\rangle$ and $|-\rangle \otimes |1\rangle$ at the angular
frequency $2E_{\rm coupl}/\hbar$. Thus after a time
$T_{0}=h/(4E_{\rm coupl})$, $|+\rangle \otimes |0\rangle$ has been
transformed into $|-\rangle \otimes |1\rangle$. At this point, a
second flux step is applied through the SQUID which drives the
system out of resonance. The dynamics is therefore "frozen" in the
superposition $|\psi(t=T_{0})\rangle=|-\rangle \otimes (\alpha
|0\rangle+ e^{i\eta}\beta |1\rangle)$ where $\eta$ is the relative
phase arising from the evolution of the initial qubit state during
time $T_0$. The full entanglement has transferred the coherent
superposition of the CPB to the SQUID, {\em i.e.}, the information
on the initial qubit state is now contained in the SQUID.

The second flux step reduces the effective critical current such
that the constant bias current is close to $I_{\rm c}$. The
barrier is therefore significantly decreased and the tunneling
rates $\Gamma _{0}$ and $\Gamma _{1}$ are drastically increased.
During the time $\Delta t$, the SQUID can escape from the well to
the non-zero voltage state by tunneling. If $\Delta t$ satisfies
$1/\Gamma _1 \ll \Delta t \ll 1/\Gamma _0$, and relaxation between
the levels is neglected, the SQUID is in its finite-voltage state
{\em if and only if the SQUID was in the state} $|1\rangle$ at
time $T_0$. In other words, the proposed measurement determines
the state of the SQUID {\em in one shot}. The escape probability
corresponds to the $|\beta|^2$ amplitude of the initial
superposition in the CPB. The lack of perfect contrast between the
escape rates from the two states and relaxation processes
introduce an intrinsic error in the proposed quantum measurement
procedure; the influence of this will be estimated later.

At $t=T_{0}+ \Delta t$, the flux is switched back to its initial
value in order to prevent further tunneling. Because the dc-SQUID
is hysteretic, the zero-voltage and finite voltage states are
stable for sufficiently long a time to perform the read-out. The
first two steps of duration $T_{0}$ and $\Delta t$ perform the
quantum measurement. The last step provides the classical read-out
measurement.

To check the feasibility of our measurement procedure, we use some typical values for
parameters of an aluminium superconducting circuit. For the CPB, we choose $E_{\rm
J}=26.2$ $\mu$eV, $C_{\rm g}=10$ aF and $C_{\rm J}=0.63$ fF, and for the SQUID,
$I_{0}=1$ $\mu$A, $I=1.1$ $\mu$A, $C_{\rm S}=1$ pF and $\Phi _{\rm DC}/\Phi
_{0}=0.277$. For $\delta \Phi (t))/\Phi _{0}=0.013$ during the first step, the
resonance condition is satisfied with $5$ levels in the well; the escape time from the
level $|1\rangle$ is much longer than 1 ms. An additional increase of flux $\delta
\Phi(t)/\Phi_{0}$ by the same amount is enough to change $\hbar \omega _{\rm p}$ by
$4$ $\mu$eV ($\gg E_{\rm coupl}$) and the system is driven out of resonance. The
escape time of level $|1\rangle$ drops to about 1 ns. Finally we choose $C_{\rm
c}=0.1$ fF yielding $E_{\rm coupl}=0.21$ $\mu$eV. Using the measuring times
$T_{0}=4.5$ ns and $\Delta t=$ 5 ns, the one-shot quantum measurement can be
performed.

\begin{figure}
\resizebox{.38\textwidth}{!}{\includegraphics{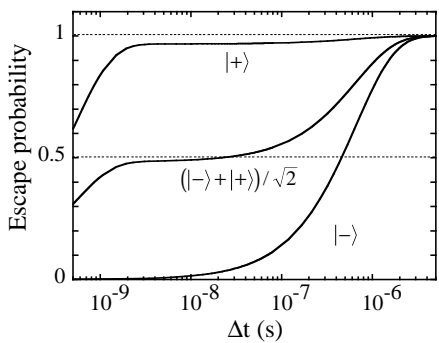}}
\caption{\label{Escape} Escape probability for three different
initial states : $|-\rangle$ , $|+\rangle$ and $ (|-\rangle
+|+\rangle)/\sqrt{2}$ for $Q=500$. The parameters for the circuit
used in the calculation are given in the text.}
\end{figure}

We now turn to the influence of relaxation in the SQUID on the
measurement procedure. At low temperature the rate $\Gamma _{\rm
R}$ of relaxation between adjacent levels down due to interaction
with the environment is dominant. Assuming $\Gamma _1 \gg \Gamma
_0,\Gamma _R$, the escape probability at $t=T_{0}+ \Delta t$ is
given in the lowest order by
\begin{eqnarray}
&& P^{e}_{\alpha |0\rangle+\beta |1\rangle}=|\beta |^2  + |\alpha
|^2 [1-e^{-\Gamma _{0} \Delta t}] \nonumber \\ && - |\beta |^2 [
e^{-\Gamma _{1} \Delta t}+(\Gamma _{\rm R}/\Gamma _{1})
(e^{-\Gamma _{0} \Delta t} - e^{-\Gamma _{1} \Delta t})] .
\label{EscapeProba}
\end{eqnarray}
Neglecting the influence of relaxation in Eq.~(\ref{EscapeProba}), and assuming an
infinite contrast between $\Gamma _0$ and $\Gamma _1$, the escape probability gives
the $|\beta|^2$ amplitude of the initial superposition in the CPB. The finite contrast
between the two states introduces an intrinsic error in the proposed quantum
measurement procedure which is about 0.8\%. Taking into account the relaxation
processes, the escape probability versus the pulse duration $\Delta t$ is plotted in
Fig.~\ref{Escape} using the experimental parameters listed above for three different
initial states : the two states $|-\rangle$ and $|+\rangle$ and the coherent
superposition $ (|-\rangle+|+\rangle)/\sqrt{2}$. The relaxation rate is defined as
$\Gamma _R=\omega _{\rm p}/Q$ where the Q-factor was chosen to be $Q=500$.
The measurement of the escape during $\Delta t=$ 5 ns is a direct measurement of the
$|\beta|^2$ amplitude. For $Q=500$, the total error of the one-shot measurement is
less than 4\%. The presented values of 0.8\% and 4\% error are overestimates because the semiclassical
approximation of the tunnelling rate slightly underestimates tunnelling from the excited state
in particular.

It is important to maintain coherence of the qubit. Specifically, we will discuss the
backaction, {\em i.e.}, dephasing of the qubit, due to the measuring circuit formed by
the SQUID and its electrical environment. To analyze this we need to study the
amplitude of the fluctuations of the phase across the CPB. The measurement environment
consists of a parallel LRC circuit, where $L_{\rm S}=\sqrt{2}\frac{\hbar}{2eI_{\rm
c}}(1-I/I_{\rm c})^{-1/2}$ is the Josephson inductance of the SQUID and $R$ describes
the dissipative part of the circuit. It is straightforward to obtain the phase
fluctuations~\cite{Ingold92}, $\langle (\Delta \phi)^2\rangle \equiv \langle
[\phi(t)-\phi(0)]^2\rangle$, where $\phi(t)$ is the phase difference across the small
junction in the CPB at time $t$, using the fluctuation-dissipation theorem. These
phase fluctuations are determined by the real part of the impedance seen by the small
junction. If the environment would be purely dissipative (resistance $R$),
we would obtain fluctuations which
diverge in time. On the contrary, the inductively shunted circuit, realized by the
SQUID, protects the qubit: asymptotically ($t \rightarrow \infty$), the expectation
value of phase fluctuations levels off to~\cite{Hekking02}
\begin{equation} \label{eq11}
\langle (\Delta \phi)^2\rangle _{\infty} \simeq \frac{2\pi}{R_{\rm
Q}} \sqrt{\frac{L_{\rm S}}{C_{\rm S,eff}}}\left(\frac {C_{\rm
c}+C_{\rm g}}{C_{\rm J}+C_{\rm c}+C_{\rm g}} \right)^2 .
\end{equation}
Here $R_{\rm Q}\equiv h/4e^2$ is the resistance quantum, and we
assumed low temperatures $k_BT \ll \hbar/(L_{\rm S}C_{\rm
S,eff})$. According to Eq.~(\ref{eq11}) the inductance, $L_{\rm
S}$, provides a protection against dephasing: the dissipative part
of the environment does not affect the value of $\langle (\Delta
\phi)^2\rangle _{\infty}$. Using the typical experimental values
given above we obtain $\sqrt{\langle (\Delta \phi)^2\rangle}
\simeq$ 0.02, which is much smaller than $\pi$, thereby
demonstrating the weakness of the residual phase fluctuations. The
low temperature condition is verified if $T \ll$ 1 K. As we have
shown, the SQUID provides protection of the qubit from the
decoherence induced by the environment. Finally, the proposed
measurement can be realized at the optimum point, where the
background charge induced decoherence is largely suppressed.

In summary, we have shown theoretically that a two junction SQUID
can perform a single shot quantum measurement of a Josephson
charge qubit. We discussed the limits of this detector posed by
the finite contrast in measuring the quantum state of the SQUID,
the finite quality factor of the SQUID, and coupling of the qubit
to environment noise.

We thank F. Faure, Ph. Lafarge, L. L\'evy, and P. Maia Neto for
useful discussions. This work was supported by the French ACI
program. JP acknowledges financial support from CNRS and Joseph
Fourier University; FH is supported by Institut Universitaire de
France.

\end{document}